\documentstyle[epsfig,mncite]{mn}
\ifCUPmtlplainloaded \else

\fi
\title[Echoes in X-ray Binaries]
{Echoes in X-ray Binaries}
\author[K. O'Brien et al.]
	{K. O'Brien,$^{1,2}$ Keith Horne,$^1$ R. I. Hynes,$^{3,4}$ W. Chen,$^4$ C. A. Haswell$^4$ and \cr M. D. Still,$^{1,5,}\thanks{also Universities Space Research Association}$ \\
$^1$ School of Physics and Astronomy, University of St. Andrews, St. Andrews KY16 9SS \\ 
        $^2$ Astronomical Institute ``Anton Pannekoek'', University of Amsterdam, 1098-SJ Amsterdam, The Netherlands \\
$^3$ Department of Physics and Astronomy, University of Southampton, Highfield, Southampton, SO17 1BJ \\
$^4$ Department of Physics and Astronomy, Open University, Walton Hall, Milton Keynes, MK7 6AA \\
$^5$ Goddard Space Flight Center, Greenbelt, MD 20771, USA}

\pagerange{\pageref{firstpage}--\pageref{lastpage}}

\begin{document} 
%
%
\newcommand{\novasco}{GRO\,J1655-40} 
\newcommand{\novaper}{GRO\,J0422+32}
\newcommand{\novamus}{X-ray Nova Muscae 1991} 
\newcommand{\sco}{Scorpius~X-1} 
\newcommand{\Her}{Hercules~X-1}
\newcommand{\her}{Hercules~X-1}
\newcommand{\cyg}{Cygnus~X-2}
%
%
\newcommand{\HST} {\textit{HST}}
\newcommand{\xte} {\textit{RXTE}} 
\newcommand{\rxte}{\textit{RXTE}}
\newcommand{\XTE} {\textit{RXTE}} 
\newcommand{\RXTE}{\textit{RXTE}}
\newcommand{\GRO} {\textit{GRO}} 
%
%
\newcommand{\HI} {H\,\textsc{i}} 
\newcommand{\HII} {H\,\textsc{ii}} 
\newcommand{\HeI} {He\,\textsc{i}} 
\newcommand{\HeII} {He\,\textsc{ii}}
\newcommand{\HeIII}{He\,\textsc{iii}} 
%
%
\newcommand{\EBV}{E(B-V)} 
\newcommand{\Rv} {R_{\rm V}} 
\newcommand{\Av} {A_{\rm V}} 
%
%
\newcommand{\lam} {$\lambda$}
\newcommand{\lamlam}{$\lambda\lambda$} 
%
%
\newcommand{\comm}[1]{\textit{[#1]}} 
\newcommand{\etal}{{\it et al}.\ } 
\newcommand{\mdot}{\dot{M}}
\newcommand{\msolar}{M_{\sun}}
\newcommand{\degree}{$^{\circ}$}
\label{firstpage}
\maketitle 
%
%
\begin{abstract} 
We present a method of analysing the correlated X-ray and optical/UV variability in X-ray binaries, using the observed time delays
between the X-ray driving lightcurves and their reprocessed optical echoes. This allows us to determine the distribution of reprocessing sites within the binary. We model the time-delay transfer functions by simulating the distribution of reprocessing regions, using geometrical and binary parameters. We construct best-fit
time-delay transfer functions, showing the regions in the binary
responsible for the reprocessing of X-rays. We have applied this model to observations of the Soft X-ray Transient, \novasco. We find the optical variability lags the X-ray variability with a mean time delay of 19.3$\pm{2.2}$ seconds. This means that the outer regions of the accretion disc are the dominant reprocessing site in this system. On fitting the data to a simple geometric model, we derive a best-fit disk half-opening angle of 13.5$^{+2.1}_{-2.8}$ degrees, which is similar to that observed after the previous outburst by \scite{orosz97}. This disk thickening has the effect of almost entirely shielding the companion star from irradiation at this stage of the outburst. 
\end{abstract}
%
%
%
%
\section{Introduction}

X-ray binaries (XRBs) are close binaries that contain a relatively
un-evolved donor star and a neutron star or black hole that is thought to
be accreting material through Roche-lobe overflow. Material passing
through the inner Lagrangian point moves along a ballistic trajectory
until impacting onto the outer regions of an accretion disk. This material
spirals through the disk, losing angular momentum, until it accretes onto
the central compact object. X-rays are emitted from inner disk regions via
thermal bremsstrahlung with an effective temperature \(\sim10^{8}K\). The
X-ray flux depends on the mass transfer rate, which in turn depends on the
structure of the disk and it's ability to transport angular momentum. In
XRBs the structure of the accretion disc is governed by irradiation.

Much of the optical emission in XRBs arises from reprocessing of X-rays by
material in regions around the central compact object. The disk is highly
ionized and out-shines the donor star. Light travel times within the
system are of order 10s of seconds. Optical variability may thus be
delayed in time relative to the X-ray driving variability by an amount
characteristic of the position of the reprocessing region in the binary
and the geometry of the binary. The optical emission may be modeled as a
convolution of the lightcurve of the X-ray emission with a time-delay
transfer function. 

This time delay is the basis of an indirect imaging technique, known as
echo tomography, to probe the structure of accretion flows on scales that
cannot be imaged directly, even with current interferometric techniques. 
Echo mapping has already been developed to interpret lightcurves of Active
Galactic Nuclei (AGN), where time delays are used to resolve photoionized
emission-line regions near the compact variable source of ionizing
radiation in the nucleus. In AGN the timescale of detectable variations is
days to weeks, giving a resolution in the transfer functions of 1-10 light
days (\scite{krolik91}; \scite{kdh91}. In XRBs the binary separation is
$\sim$ light seconds rather than light days, requiring high-speed optical
and X-ray lightcurves to probe the components of the binary in detail. The
detectable X-ray and optical variations in the lightcurves of such systems
are also suitably fast. Extreme examples of this rapid variability are
X-ray bursts, for example from \cyg~\cite{kuulkers1995}, where the X-ray
flux can increase by \(>\) 50$\%$ with a rise time of 2-3 seconds and a
duration of $\sim$5 seconds. Time-delayed optical bursts have been seen
clearly in the object 4U/MXB~1636-53 (\scite{pedersen82}; 
\scite{lawrence83}; \scite{matsouka84}). Pulsed X-ray emission from \Her\
produces faint optical pulsations that are thought to be echoes from the
irradiated companion star \cite{middleditch76}.

Recently, \scite{hynes98echo} found correlated time-delayed X-ray and UV
variability in the lightcurves of the Soft X-ray transient \novasco, using
\XTE\ and \HST. The data, centred around binary phase 0.4, shows a mean
time-delay of 14 seconds and an RMS delay of 10.5 seconds. The mean time
delay is consistent with reprocessing in the outer regions of the
accretion disk, while the relatively large range of delays suggests that
there is relatively little companion star reprocessing. This may occur if
a thick outer accretion disk shields the companion star from the X-ray
source.

In this paper we present a simple geometric model for the time-delay
transfer functions of XRBs, using a synthetic binary code. We analyse
correlated X-ray and UV variability in \novasco, using our
computed transfer functions, to constrain the size, thickness and
geometric shape of the accretion disk. 

Section~\ref{optlightcurves} describes the features seen in typical optical lightcurves of LMXBs and how these affect our analysis. Section~\ref{reprocessing} describes the reprocessing of X-rays in XRBs in more detail, including the reasons for the time delays between the X-ray and optical variability.
Section~\ref{code} describes the model created to describe the time delays
found, while Section~\ref{gauss} describes an alternative method that uses
Gaussian transfer functions. Section \ref{novasco} shows the
results of this analysis. This is
followed by a discussion of the results in Section~\ref{discuss}.  

%
%
\section{Optical lightcurves}
\label{optlightcurves}
The lightcurves of X-ray binaries contain many temporal and spectral features (see \scite{vanparadijs95} for a review of many of these), some periodic and others purely random or quasi-periodic in nature (See \scite{vdk2000} for a review). The periodic features, such as eclipses and the minima and maxima of lightcurves are due to the orbital motion of the components of the binary around the centre of mass of the system. These periodic features are relatively easy to model and much success has been made of such models in determining the geometric parameters of binaries using the information in multi-colour lightcurves. Refinements have been made to include effects such as limb darkening, caused by an decreasing source function with projected radius from the stellar core, and gravity darkening, which causes a change in temperature of the distorted shape of the roche-lobe filling star (eg. \scite{orosz97}).

In the case of X-ray binaries the intense X-ray flux from the regions surrounding the compact object also affect both the periodic and non-periodic features of the observed optical lightcurves. In the standard model of reprocessing, X-rays are emitted by material in the deep potential well of the compact object. These photoionize and heat the surrounding regions of gas, which later recombine and cool, producing lower energy photons. Hard X-rays that penetrate below the photosphere emerge as continuum photons, with an energy distribution characteristic of the temperature of the photosphere. Soft X-rays that are absorbed above the photosphere emerge as emission line photons from a temperature inversion layer near the surface of the disk. Such irradiation will have the effect of changing the overall form of the orbital lightcurve as the aspect of the hot, irradiated regions changes with the binary orbit, this effect is very noticeable with the changing irradiation of the inner face of the companion star in \her, during the 35-day precession phase of the tilted accretion disc \cite{boynton73}. While these features are all vital in interpreting the long timescale orbital lightcurves of X-ray binaries, many of them are not important in this work, which deals with a very small range of binary phases (less than 1\%). 

The short duration of the observations is also important when considering the effects of other transient features,  such as star spots, whose lifetime is much longer than the duration of our observation and whilst it will undoubtedly make a small effect on our model fits, the uncertainties involved in including them far outweigh any benefits to the model from including them. 

In our model, we have assumed that all the optical variability is in fact caused by the reprocessing of X-ray irradiation. Furthermore, we have assumed that the thermal component will dominate the reprocessed emission in our broadband observations. These assumptions are clearly not strictly true, observations of cataclysmic variables and other interacting binaries, where irradiation is no longer the dominant source of optical emission, also show non-periodic optical variability superimposed on the periodic orbital lightcurves (eg. the Nova-like AE Aquarii, \scite{welsh93}). However, while such non-correlated features will introduce noise into out analysis, thus increasing the absolute value of the badness-of-fit of our models, they will have little effect on the relative values, which have contributions from all points in the lightcurve. Similarly, the optically thin emission is only a small fraction of the total broadband emission and for this reason is assumed to vary in phase with the thermal emission.
%
%
\section{Reprocessing of X-rays} 
\label{reprocessing} 
The reprocessed, optical emission seen by a distant observer is delayed in time of arrival relative to the X-rays by two mechanisms. The first is a finite
reprocessing time for the X-ray photons and the second is the light travel
times between the X-ray source and the reprocessing sites within the
binary system.

\subsection{reprocessing times}

The average reprocessing time for line photons is given by the average
recombination time, \cite{hummer63}
\begin{equation}
\frac{\tau_{\mbox{rec}}}{\mbox{s}} \sim \left(\frac{n_e}{10^{13}\,\mbox{cm}^{-3}}\right)^{-1}.
\end{equation}

In the accretion disk the high electron density, $n_{e} \sim
10^{15}$\,cm$^{-3}$, ensures that the time-scale for reprocessing of line
photons is short compared to the overall time delay. 

The continuum photons which scatter from deeper within the accretion disk
undergo a `random walk' before escaping through the photosphere, leading
to a longer reprocessing time. The exact determination of this continuum
reprocessing time is complicated, requiring detailed model atmosphere
calculations, which is outside the scope of this paper. However, the short
time delays between X-ray and optical bursts \cite{pedersen82} and the
observations of 1.24 second optical pulsations from \her\
\cite{chester79}, imply that a significant fraction of the reprocessed
optical photons emerge from the reprocessing site within $\sim0.6~$seconds
of the absorption of the incident X-ray photons. This delay is smaller
than the measured uncertainty in the mean delay for the systems and so we
have treated the reprocessing as instantaneous. We have also treated the
reprocessing of X-rays as ``passive'' reprocessing, where the absorbed
X-rays do not affect the structure of the material in the binary.  

\subsection{light travel times}

The light travel times arise from the time of flight differences for
photons that are observed directly and those that are reprocessed and
re-emitted before traveling to the observer. These delays can be up to
twice the binary separation, obtained from Kepler's third law, 
\begin{equation}
\frac{a}{c} = 9.76\mbox{s} \left(\frac{M_{x}+M_{c}}{\msolar}\right)^{\frac{1}{3}}  \left(\frac{P}{\mbox{days}} \right)^{\frac{2}{3}}
\end{equation}
where $a$ is the binary separation, $M_{x}$ and $M_{c}$ are the masses of
the compact object and donor star, $P$ is the orbital period. In LMXBs the
binary separation is of the order of several light seconds.

The time delay $\tau$ at binary phase $\phi$ for a reprocessing site with
cylindrical coordinates (R, $\theta$, Z) is
\begin{equation}
\label{delayeqn}
\tau(\b{x},\phi) = \frac{\sqrt{R^2 + Z^2}}{c}(1+\sin{i}\cos(\phi-\theta)) -\frac{Z}{c}\cos{i}
\end{equation}
where $\it{i}$ is the inclination of the system and $\it{c}$ is the speed
of light. This can also be expressed using the position vector, $\b{x}$
and the unit vector, $\b{e}(\phi)$, pointing toward the earth,
\begin{equation}
\tau(\b{x},\phi) = \frac{|\b{x}|-\b{e}(\phi).\b{x}}{c}
\end{equation}


The X-ray driving lightcurve, $f_{x}(t)$, is described as the sum of a
constant and a variable component,
\begin{equation}
f_{x}(t) = \overline{f_{x}} + \Delta f_{x}(t).
\end{equation}
The reprocessed lightcurve, $f_{\nu}(t)$, is similarly divided into two
components. The relationship between the two is given by
\begin{equation}
f_{\nu}(t) = \overline{f_{\nu}} + \int \Psi_{\nu}\left(\lambda, \tau, \phi \right) \left( f_{x}\left(t-\tau \right) - \overline{f_x} \right) d\,\tau
\end{equation}

where $\Psi_{\nu} \left( \lambda, \tau, \phi \right)$ is the time delay
transfer function. This transfer function is the strength of the
reprocessed variability delayed by $\tau$ relative to the X-ray
variability. 

%


The dynamic response function is found by considering how a change in X-ray flux drives a change in the reprocessed flux. We can define the dynamic time delay transfer function to be,
\begin{eqnarray}
\Psi_{\nu}\left(\lambda,\tau,\phi\right) & = & \int \left[ \frac{\delta I_{\nu}\left(\lambda,\b{x}, \Delta f_{x}(t-\tau) \right)}{\delta f_{x}(t-\tau)}\right]. \nonumber \\
&& d \Omega(\b{x},\phi) \, . \, \delta (\tau - \tau(\b{x},\phi))
\label{dynamic}
\end{eqnarray} 
where $\tau(\b{x},\phi)$ is the geometric time delay of a reprocessing
site at position $\b{x}$, see Equation~\ref{delayeqn}. In the next section
we describe the model X-ray binary code we have been using and how we have
used this to find the reprocessed flux.
%
%
\section{Model X-ray Binary code}
\label{code}

We have developed a code to model time delay transfer functions based on
determining the contributions from different regions in the binary. In
this section we describe the models used to construct the individual
regions of the binary; the donor star, the accretion stream and the
accretion disk. The code uses distances scaled to the binary separation in
a right-handed Cartesian coordinate system corotating with the binary. The
X-direction is along the line of centres for the binary, the Y-direction
is perpendicular to this in the orbital plane of the binary, so that the
X-ray source is at (0,0,0) and the centre of mass of the donor star is at
(1,0,0). Each surface element is a triangles, characterized by its area
$dA$, orientation $\b{n}$, position $\b{x}$ and temperature $T$.

We calculate the total monochromatic flux by summing up contributions from
all visible elements. The area $dA$ and the normal vector $\b{n}$ for the
triangular panel are calculated and then the projected area of the panel
is calculated using the projected earth vector \b{e}($\phi$,$\it{i}$). The
effects of occultations by regions in the binary are also considered.
Therefore the observability, $O(\b{x},\phi)$, is given by,
\begin{equation}
O(\b{x},\phi) = \b{dA}(\b{x}).\b{e}(\phi)
\end{equation}
which is the foreshortened area of pixel $\b{x}$, observable at phase
$\phi$. This is related to the solid angle of the observed pixel,
$d\Omega$ by the relation
\begin{equation}
\label{viseqn}
d\Omega(\b{x},\phi) = \frac{O(\b{x},\phi)}{D^2},
\end{equation}
where $D$ is the distance to the source. If $d\Omega(\b{x},\phi)<0$ then
the panel is not visible to the observer and does not contribute to the
flux. The monochromatic intensity is calculated using the Planck function,
\begin{equation}
\label{bbeqn}
B_{\nu}(\lambda,T)=\frac{2hc}{\lambda^{3}\left[ \exp \left( \frac{hc}{k \lambda T}\right) -1 \right]}
\end{equation}
This is scaled using the projected area of the panel $\left| O(\b{x},\phi)
\right|$, as seen by the observer. The standard linear limb-darkening law
is assumed,
\begin{equation}
\label{limbeqn}
L(u,\alpha)=\frac{1-u+u\,\cos{(\alpha)}}{1+u/3}
\end{equation}
where $u$ is the linear limb-darkening coefficient, assumed to be 0.6, and
$\alpha$ is the angle between the normal vector $\b{n}$ and the earth
vector \b{e}($\phi$,$\it{i}$),
\begin{equation}
\cos{\alpha} = \b{n}.\b{e}.
\end{equation}
%
%

The response curve for a given detector, $P\left(\lambda\right)$, is
combined with the limb-darkened Planck function to create the synthetic
reprocessed flux from each visible triangular element. The total detected
flux due to reprocessed X-rays from a single panel, $F_{\nu}(\lambda,t)$,
is given by,
\begin{equation}
F_{\nu}(\lambda,T) = \int B_{\nu}\left(\lambda,T\right) P\left(\lambda\right) L\left(u,\alpha\right) d\Omega \left( \b{x} , \phi\right) d\lambda,
\end{equation} 

where $L\left(u,\alpha\right)$ and $d\Omega\left( \b{x} , \phi\right)$ are
the expressions for the limb-darkening and the solid angle of the exposed
panel, see equations~\ref{limbeqn} and \ref{viseqn} respectively. In the next sections we describe the geometric model used to calculate the time delay for a given panel and the effects of irradiation which are used to determine the contribution of each panel to the final transfer function.
\subsection{Donor Star}
\label{star}

The donor star is modeled assuming it fills its critical Roche potential,
so that mass transfer occurs via Roche lobe overflow through the inner
Lagrangian point. Optically thick panels are placed over the surface of
the Roche potential. The panels are triangular so that the curved surfaces
of the binary are mapped more accurately than is possible using 4-sided
shapes \cite{rutten94}. These panels are equally spaced in longitude and
latitude across the surface of the star. In order to correctly specify the temperature of each panel on the face of the tidally distorted star, one must take into account the degree of gravity darkening. Using von Zeipel's theorem \cite{zeipel24}, for the relationship between the local gravity and the local emergent flux one finds that the relationship between the local temperature, $T_{e}$, and the gravity, $g$, is
\begin{equation}
T^{4}_{e}(\b{x}) \propto g(\b{x}).
\end{equation}
As a consequence the temperature at any point on the star is given by 
\begin{equation}
\frac{T\left(\b{x}\right)}{T_{pole}} = \left[\frac{g\left(\b{x}\right)}{g_{pole}}\right]^{\beta},
\end{equation}
where $T_{pole}$ and $g_{pole}$ are the temperature and gravity of the pole of the star. The ``gravity darkening exponent'' $\beta$ is 0.25 for stars with fully radiative envelopes (\pcite{zeipel24}; as is the case in our models) and 0.08 for stars with fully convective envelopes \cite{lucy66}. $T_{pole}$ is taken to be the effective temperature of a field star with a similar spectral type to the donor star.

\subsection{Accretion stream}
\label{stream}

The accretion stream is modeled by following the ballistic trajectories of
4 test particles. The thickness (w) of the stream defines the initial
positions of the test particles. These test particles determine the
`width' of the stream (Its deviation from the line of centres of the
binary, in the plane of the binary, or y-direction) and the `height' of
the stream (it's extent in the direction normal to the plane of the
binary, the z-direction), assuming the stream is symmetric about the x-y
plane.

The particles start at the L1 point with a small velocity in the direction
of the compact object (-v,0,0), from positions (R(L1),0,0), (R(L1),w,0),
(R(L1),-w,0) and (R(L1),0,w). The trajectory is cut into discrete steps,
with the step size as a parameter of the code. The velocity and position
of each particle are determined from the Roche potential after each step. 
The stream is curtailed when one of two criteria are reached; (1) the
stream has collapsed vertically, or (2) the core trajectory is moving
outwards, ie. the stream has passed the compact object without collapsing
vertically. 

The unirradiated accretion stream is assumed to have a constant
temperature $T_{s}$ along it's length and the effects of irradiation are
considered in the same way as those of the donor star in Section~\ref{irrad}. 
\subsection{Accretion disk}
\label{disk}
The disk thickness is assumed to increase with radius from 0 at
$\mbox{R}=\mbox{R}_{in}$ to H$_out$ at $\mbox{R}=\mbox{R}_{out}$, with the
form,
\begin{equation}
H= R_{out} \left( \frac{H}{R}\right )_{out}\left( \frac{R-R_{in}}{R_{out}-R_{in}}\right )^{\beta},
\end{equation}
where the parameters are the inner and outer disk radii, $R_{in}$ and
$R_{out}$ in units of R(L1), the half thickness of the outer disk
$(H/R)_{out}$ and the exponent $\beta$ which describes the overall shape
of the disk. The temperature structure of the un-irradiated disk is that
of a steady state disk, in the absence of irradiation,
\begin{equation}
\label{disktemp}
T_{disk}(R)=T_{out}\left( \frac{R}{R_{out}}\right)^{-\frac{3}{4}},
\end{equation}
where $T_{out}$ is the temperature at the outer disk and $T_{in}$ is the
temperature of the inner disk. 

The disk is divided radially and azimuthally, into $N_R$ and $N_{\theta}$
sections. The monochromatic intensity is again calculated using the Planck
function, corrected for limb-darkening using a linear limb-darkening law,
with a constant coefficient. The intensity is again scaled using the
projected area of the panel, for the given values of binary phase $\phi$
and inclination $\it{i}$.

\subsection{Irradiation model}
\label{irrad}
The effective temperature of a region at a distance R from the X-ray
source, assumed in our model to be a point source located at the centre of
the accretion disk, is found from the accretion luminosity for a typical
LMXB,
\begin{equation}
\label{tx}
T_{x}^{4} = \frac{L_{x} (1-A)}{4\pi \sigma R^2}
\end{equation}
and 
\begin{equation}
L_x = \eta \frac{ GM_{x}\mdot}{R_{ns}}
\end{equation}

where $T_{x}$ is the temperature, A is the albedo, $\eta$ is the
efficiency, $M_{x}$, the mass of the compact object, $\mdot$ the accretion
rate onto the compact object, $R_{ns}$ is the size of the compact object
and R is the distance between the compact object and the irradiated
element. This is normalised using the binary separation, a, the distance
between the centres of mass of the stars, as is the coordinate system for
the binary. In the case of \sco, this gives $T_x \sim 10^5 K$ for a 1.4
$\msolar$ neutron star ($R_{NS} \sim 10 km$) accreting $10^{-9}
\dot{M}_{\sun} yr^{-1}$, with an efficiency $\eta =0.1$, an albedo of 0.5,
at a distance equal to the binary separation of 3.4 x $10^{11}$ cm. 

The irradiation of the binary takes place in three stages. The first stage
is to calculate the temperature structure of the binary in the absence of
any irradiation. This is done with characteristic temperatures for the
donor star (from it's spectral type) and the accretion stream and disk. 
The temperature structure of the disk is assumed to that for an
unirradiated disk as given in equation~\ref{disktemp}. The surface
elements of the binary exposed to X-rays are determined by projecting the
binary surfaces onto a spherical polar representation of the sky, as it
appears from the X-ray source. Each triangular element is mapped to the
sky starting with the one furthest from the source and ending with the
triangle closest. Those elements remaining visible and unocculted on the
sky map are irradiated. The change in effective temperature of an element
is scaled by the projected area with respect to the X-ray source at a
distance R from the source. Hence the temperature after irradiation is
given by,

\begin{equation}
\label{tempx}
T^{4} = T_{x}^{4} \cos\theta \left( \frac{a}{R} \right)^{2} + T_{eff}^{4}
\end{equation}
where $T$ is the temperature of the panel, $\theta_x$ the angle between
the line of sight from the central source and the normal to the surface of
the element and $T_{eff}$ is the unirradiated effective temperature of the
panel. 

Since we are interested in finding the correlation between the variable
component of the X-ray and reprocessed fluxes, we split the X-ray flux
into constant and time-dependent components. These components of the flux
are converted into components of temperature, 
\begin{equation}
T_{x}(t) = \overline{T_{x}} + \Delta T_{x}(t)
\end{equation}
where 
\begin{equation}
\frac{\Delta{f_{\nu}(t)}}{\overline{f_{\nu}}} =  \frac{4 \Delta T_{x}(t)}{\overline{T_x}}.  
\end{equation}

The second stage is to irradiate the binary with the constant component of
the X-ray flux. This component of the X-ray flux is equated to the mean
effective temperature of the X-ray source, as given in
equation~\ref{tempx}, where $T_x \equiv \overline{T_{x}}$. The third and
final stage is to repeat stage two with $T_x \equiv {T(t)_{x}}$, which
represents irradiating the binary with a time varying component. The
difference between stages two and three represents the temperature change
of the elements due to the time varying component of the X-ray flux alone,
$\Delta f_{\nu}(t)$. 

Thus the response of a panel to the variable component of the irradiating
X-ray flux is given by,
\begin{eqnarray}
I_{\nu}\left(\lambda,\b{x},\Delta f_{x}(t)\right) & = & \int \left[ B_{\nu}\left(\lambda,T_{x}(t)\right) - B_{\nu} \left( \lambda, \overline{T_{x} } \right) \right] . \nonumber \\
&& P\left(\lambda\right) I\left(u,\alpha\right) d\Omega \left( \b{x} , \phi\right) d\lambda.
\end{eqnarray}
This response is substituted into the expression for the dynamic response
given in equation~\ref{dynamic}. 

\subsection{Transfer functions}
\label{delays}

In order to transform the reprocessed flux into a time delay transfer
function, we define iso-delay surfaces. These surfaces are nested
paraboloids around the line of sight to the X-ray source, defined by the
earth vector, \b{e}($\phi$,$\it{i}$). The parabolic surfaces have a mean
time delay $\tau$ and a width $\delta\tau$. The mean time delay $\tau$ in
seconds between the directly observed X-ray flux from the central source
and the reprocessed signal from a point with cylindrical coordinates
(R,$\theta$,Z) is
\begin{equation}
\tau = \frac{\sqrt{R^2 + Z^2}}{c}(1+\sin{i}\cos(\phi-\theta)) -\frac{Z}{c}\cos{i}
\end{equation}
where $\it{i}$ is the inclination of the system, $\phi$ is the binary
phase and $\theta$ is the angle from the line of centres of the binary. 

\begin{figure*}
\begin{center}
\epsfig{width=6.0in,angle=0,file=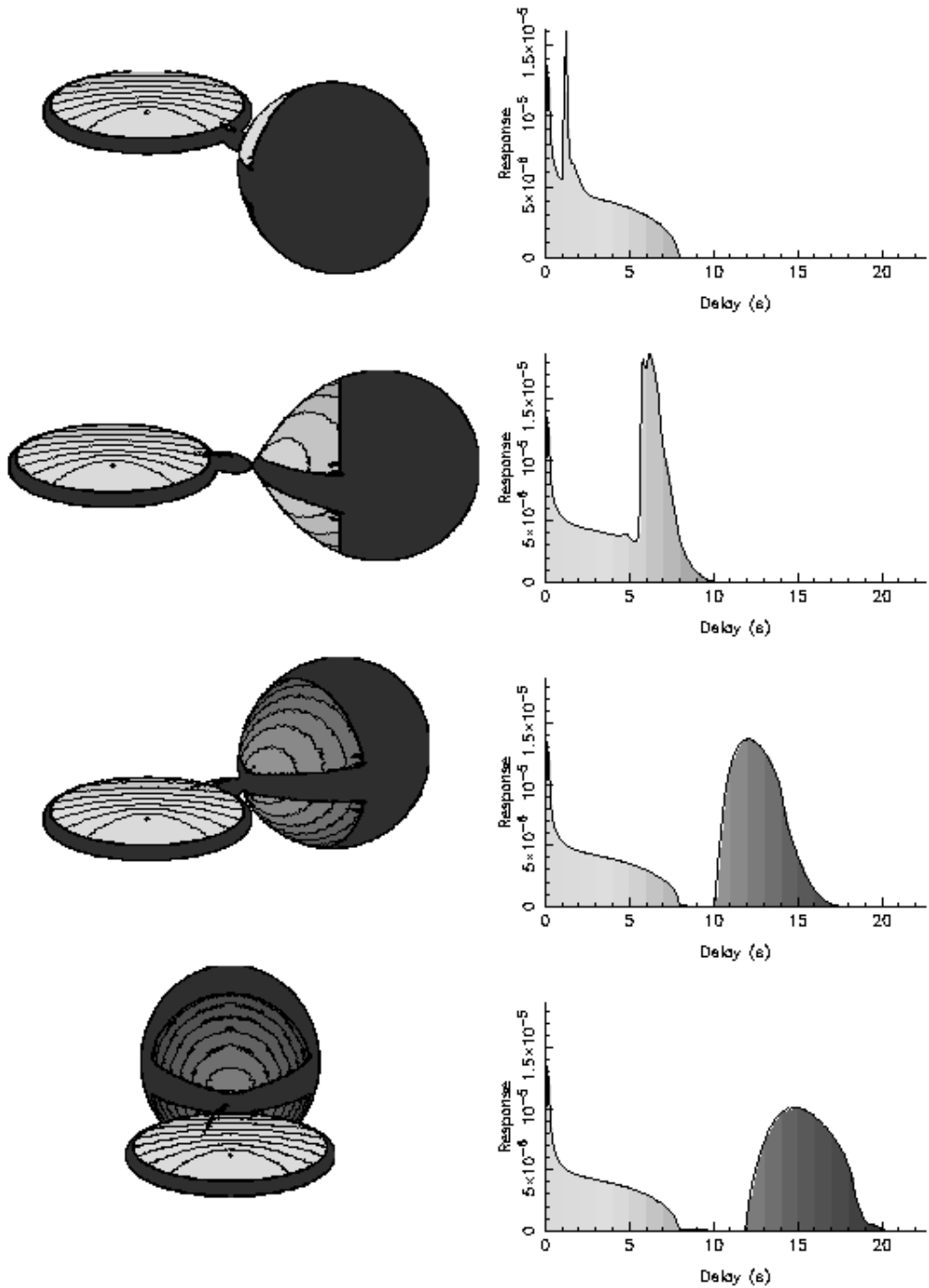}
\caption{Left, model X-ray binaries, based on typical binary parameters for an LMXB, showing iso-delay surfaces projected onto the irradiated surfaces of the binary. Right, the associated time delay transfer functions, showing the relative contributions from the regions highlighted in the model X-ray binaries.}
\label{isoplot}
\end{center}
\end{figure*}

We consider models with reprocessing taking place in the accretion disk, accretion stream and companion star. Each of these regions makes a contribution to the transfer function for the system, see Figure~\ref{isoplot} for a diagram of the results of our code and the calculated transfer functions. 

\begin{figure}
\begin{center}
\epsfig{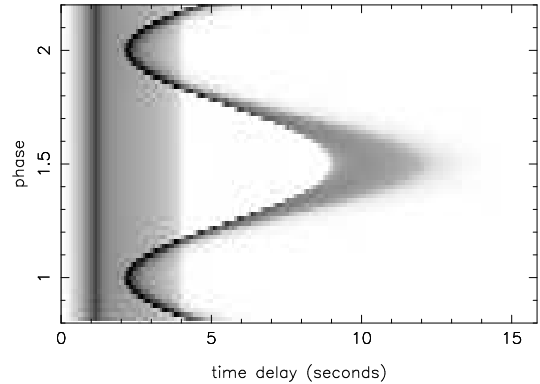}
\caption{A plot of time-delay transfer functions as a function of binary
phase, based on the typical binary parameters for an LMXB. The accretion disk has constant time delays in the
region 0-4 seconds, whereas the time delays from the companion star are
seen to vary sinusoidally with binary phase between 2 and 10 seconds.}
\label{scoechophase}
\end{center}
\end{figure}

The phase dependence of $\tau$ allows us to create time-delay transfer functions as a function of binary phase, allowing us to produce phase-delay diagrams for the system, see figure~\ref{scoechophase}. X-rays reprocessed at the companion star have a time delay that varies sinusoidally in phase
with semi-amplitude $(a/c)\sin{i}$ around a mean value $a/c$. X-rays
reprocessed by a circular disk appear as a phase independent contribution
to the delay distribution $\Phi(\tau,\phi)$ between the inner and outer
radii of the disk. The accretion stream shows up as a non-symmetric
contribution that varies roughly sinusoidally with the orbital motion of
the companion star and can be seen faintly in Figure~\ref{scoechophase} as
the contribution near superior conjunction between the outer rim of the
disk and the inner-face of the companion star.

The relative intensities of the contributions, represented by the area
under the transfer function, constrain the geometric parameters of the
system, especially the contribution from the accretion disk, which is
probably the most important region for reprocessing of X-rays in
interacting binaries.

\begin{figure}
\begin{center}
\epsfig{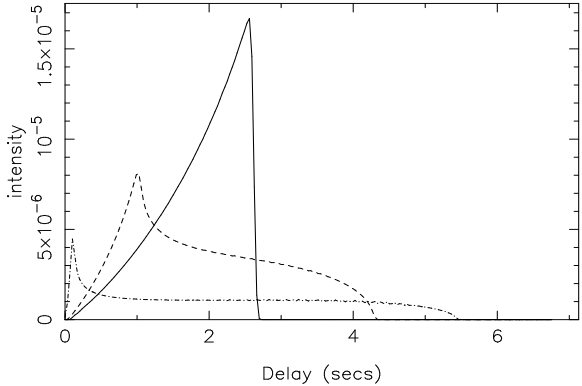}
\caption{Model time-delay transfer functions for an accretion disk as a function of binary inclination. The solid line is for a binary inclination of 0\degree, the dashed for 35\degree and the dot-dash for 70\degree}
\label{isoplotinc}
\end{center}
\end{figure}

Thus the shape of the iso-delay surfaces when projected onto the plane of
the binary depends on the inclination of the system. For a face-on disk
($\it{i} = 0^{\circ}$) the projected iso-delay surfaces are simple circles
on the disk. As the inclination of the system increases these surfaces become elongated along
the line of sight to the X-ray source, until finally they form parabolae
if viewed from edge-on ($\it{i} = 90^{\circ}$). This inclination
dependence is shown in the transfer functions in Figure~\ref{isoplotinc}.

The peak in these transfer functions is also a function of inclination,
with the peak occurring with the time delay of the largest iso-delay
surface that is contained entirely within the disk,
\begin{equation}
\tau =  \frac{R}{c}(1-\sin{(i-\delta)}),
\end{equation}
where $\delta$ is the opening angle at the edge of the disk, $\delta =
\tan{^{-1} (H/R)}$. 

\begin{figure} 
\begin{center}
\epsfig{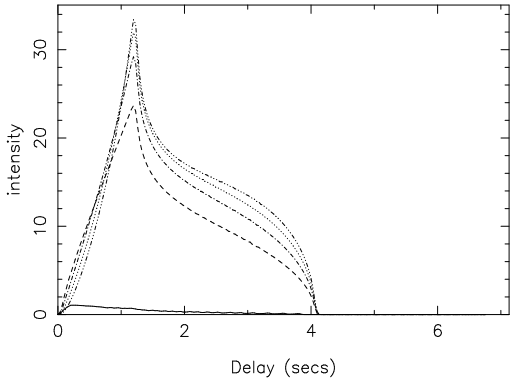}
\caption{Model time-delay transfer functions as a function of $\beta$, the
disk exponent for $\beta$= 1.01(solid line), 1.5(dashed line), 2(dot-dash
line), 2.5(dotted line) and 3(3 dots-dash line)} 
\label{isoplotbeta}
\end{center} 
\end{figure}

The disk shape, characterized by the exponent $\beta$, also affects
the position and shape of the peak in the transfer function. The
contribution is scaled with the projected area of the surface element,
therefore as $\beta$ increases the peak should move to longer time delay,
as the edge of the disk becomes steeper and the projected area decreases.
This can be seen in figure~\ref{isoplotbeta}, where time-delay transfer
functions are plotted as a function of $\beta$ with values between 1.01
and 3. (Note: When $\beta$=1, the surface of the disk is flat, whose
thickness goes to zero at the position of the compact object, therefore
the projected area of the disk as seen from the compact object is zero,
except on the inner edge. For this reason $\beta$ is given a value just
greater than 1 to show the limit as $\beta$ tends towards the case of an
un-flared disk.)

%
%
\section{Gaussian transfer functions}
\label{gauss}

In paper I Gaussian time delay transfer functions were used to create
synthetic reprocessed lightcurves for the SXT \novasco. This is a
relatively simple form of transfer function as it assumes nothing about
the geometry of the system. These transfer functions have the form,
\begin{equation}
\Psi( \tau ) = \frac{\Psi}{ \sqrt{2 \pi } \Delta \tau }
        \exp \left( - \frac{1}{2} \left(
                \frac{\tau-\tau_0}{\Delta \tau } \right)^{2} \right).
\end{equation}

Thus the transfer functions have three parameters, the mean time delay
$\tau_0$, it's variance ~$\Delta\tau$ and the strength of the response
$\Psi$. These transfer functions were convolved with the X-ray driving
lightcurve in the same way as the model transfer function and the badness
of fit between the synthetic and real reprocessed lightcurves calculated.

The equivalent parameters $\tau_0$ and $\Delta\tau$ can be determined for
any transfer function, $\Psi(\tau)$ by calculating the first and second
moments.
\begin{equation}
\tau_0 \equiv \frac{\int_{-\infty}^{\infty} \Psi(\tau)~\tau~d\tau}{\int_{-\infty}^{\infty} \Psi(\tau)~d\tau},
\end{equation}
\begin{equation}
\Delta \tau \equiv \left[ \frac{\int_{-\infty}^{\infty} \Psi(\tau)~(\tau-\tau_{0})^{2}~d\tau}{\int_{-\infty}^{\infty} \Psi(\tau)~d\tau}\right]^{1/2}.
\end{equation}
%
%

%
\begin{table}
\begin{center}
\begin{tabular}{lccc}
& & & GRO \\ 
Parameter & symbol & unit &  J1655-40 \\ \hline
Distance & D & kpc &  3.2\\
Period & P & days &  2.6 \\
Inclination & $\it{i}$ & deg. & 69.5 \\
Mass ratio & q &&  0.3344 \\
Primary mass & $M_x$ & $M_{\sun}$ & 7.02 \\
Binary sep. & a & $10^{11}$ cm & 12 \\
Star Temp & $T_{pole}$ & K & 6500 \\
Stream Temp & $T_{s}$ & K & 5000 \\
Outer Disk & $T_{out}$ & K & 6000 \\
Inner Disk & $T_{in}$ & K & 100000 \\
Irrad Temp & $T_x$ & K & 10$^{5}$ \\
Inner radius & $R_{in}$ & R($L_1$) & 10$^{-4}$ \\ 
Outer radius & $R_{out}$ & R($L_1$) & free \\
Disk thickness & H/R & & free \\
disk exponent & $\beta$ & & free \\ \hline
\end{tabular}
\end{center}
\caption{Parameters used in our model of X-ray Binary, the values are taken from {\protect\scite{orosz97}}}
\label{parstable}
\end{table}
%
%
\section{Observation and data reduction}
\label{novasco}

The Soft X-ray Transient (SXT) \novasco\ was discovered in 1994 July when
the Burst and Transient Source Experiment (BATSE) on \GRO\ observed it in
outburst at a level of 1.1\,Crab in the 20--200\,keV energy band
\cite{harmon95}. After a period of apparent quiescence from late 1995 to
early 1996, \novasco\ went into outburst again in late 1996 April
\cite{remillard96}, and remained active until 1997 August.  During the
early stages of this outburst we carried out a series of simultaneous
\HST\ and \XTE\ visits.  One of the primary goals of this project was to
search for correlated variability in the two wavebands.  The long-term
evolution of this outburst argued against significant reprocessing, as the
seemingly {\em anticorrelated} optical and X-ray fluxes observed
\cite{hynes98spec} are not to be expected if the optical flux is
reprocessed X-rays.  Nonetheless, significant short term correlations were
detected.  In our previous paper \cite{hynes98echo} we have analyzed these
correlations using both acausal and causal Gaussian transfer functions. We
were able to put constraints on the possible regions responsible for X-ray
reprocessing. In this paper we summarize this previous analysis and
develop it further, using our synthetic binary model to create more
physical transfer functions. 

\subsection{observations}

The \HST\ and \RXTE\ observations used in this paper were first presented,
along with similar exposures, in \shortcite{hynes98echo}. In paper I we
fitted causal and acausal Gaussian transfer functions to 4 similar
exposures in a similar analysis as used in the previous section. The
time-delay distribution based on all the observations is 14.6 $\pm$ 1.4
seconds, with a dispersion of 10.5 $\pm$ 1.9 seconds. 

\subsubsection{\HST}

\begin{figure}
\begin{center}
\epsfig{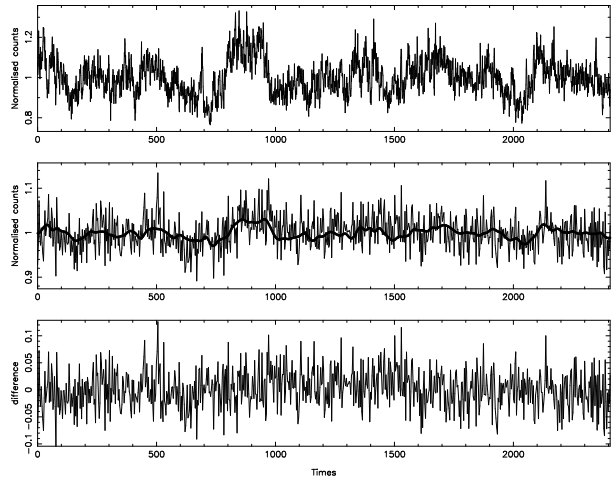}

\caption{The best-fit results for GRO J1655-40 using acausal Gaussian
transfer functions. Top panel, the normalised X-ray driving lightcurve
from RXTE. Middle panel, UV lightcurve from HST with synthetic UV
lightcurve superimposed (thick line). Bottom panel, residuals of the fit
to the UV lightcurve.}

\label{j1655light}
\end{center}
\end{figure}

The \HST\ observations are shown in figure 1(b) of paper I, along with a
detailed description of the data reduction techniques used. In this paper
we have used one of the observations, referred to as exposure 6, shown in
Figure 1(b) in \shortcite{hynes98echo} and reproduced in this paper as
figure~\ref{j1655light}. This observation took place during the June 8
1996 visits, using The Faint Object Spectrograph in RAPID mode with the
PRISM and blue detectors (PRISM/BL), covering the spectral range
$\sim2000-9000$\,\AA. The resulting lightcurve has a time resolution of
$\sim3$ seconds, the absolute time accuracy of this data is limited to
0.255 seconds. 

\subsubsection{\XTE\ }

The X-ray data was taken with the PCA onboard \RXTE\ on June 8 1996,
simultaneous with the \HST\ data. The lightcurve was created from the
standard-1 EDS mode data, using the {\rm saextrct} task in the
\textsc{ftools} software package. This mode has a maximum time-resolution
of 0.125s, but with no spectral resolution. It is desirable for
echo-mapping for the driving X-ray lightcurve to have a higher time
resolution than the reprocessed one, so that the time delay transfer
function can have a time resolution greater than the time resolution of
the reprocessed lightcurve. A lightcurve with a time resolution of 1s and
an absolute timing accuracy of about 8 $\mu$s was extracted. For a full
description of the data, see paper I. 

%
%
\subsection{results of modeling}
\subsubsection{Gaussian transfer functions}

The results for the fits to the acausal Gaussian transfer functions are
described in depth in paper I and summarized in this paper in
Table~\ref{j1655gausstable}, together with their 1-parameter 1-sigma
confidence regions, in order to compare them to the results from our
modeling of the X-ray binary. The mean and rms values for the delay in the
Gaussian transfer function represent the first and second moments expected
for the model transfer functions. The best fit Gaussian transfer function
for exposure 6 has a mean delay of 19.3s with an rms delay 10.8s. 
\scite{hynes98echo} interpreted this result as evidence for reprocessing
in the outer regions of a thick accretion disk that could be thick enough
to shield the companion star from significant irradiation. If this were
not the case and we had considerable disk and companion star reprocessing
we would expect the variance of the transfer function to be much higher. 

\begin{table}
\begin{center}
\begin{tabular}{lc}
Dataset & Exp. 6  \\ \hline
$\phi$ & 0.42 \\
N & 782 \\ \hline
$\chi^2$/(N-3) & 1.230 \\ 
$\tau_0$ & 19.3 $\pm$ 2.2 \\
$\Delta\tau$ & $10.8^{+3.7}_{-3.3}$ \\
$\Psi/10^{-3}$ & $55^{+11}_{-7}$ \\ \hline
\end{tabular}
\end{center}

\caption{Best fit values for the acausal Gaussian time delay transfer
function fitting to the X-ray and optical lightcurves for \novasco. The
label exp. 6 refers to exposure 6 in the original paper, Paper I. }
\label{j1655gausstable}
\end{table}
\subsubsection{synthetic binary transfer functions}
\label{j1655model}
\begin{table}
\begin{center}
\begin{tabular}{lc}
Dataset & Exp. 6  \\ \hline
$\phi$ & 0.42 \\
N & 782 \\ \hline
$\chi^{2}$/(N-4) & 1.230\\
$R_{out}$ & $0.65^{+0.06}_{-0.06}$ \\
H/R & $0.24^{+0.04}_{-0.05}$ \\
$\beta$ & $4.7^{+\infty}_{-2.0}$\\ \hline
\end{tabular}
\end{center}
\caption{Best fit values for the synthetic X-ray Binary model time delay
transfer function fitting to the \RXTE\ and \HST\ lightcurves for
\novasco. }
\label{j1655modeltable}
\end{table}
\begin{figure}
\begin{center}
\epsfig{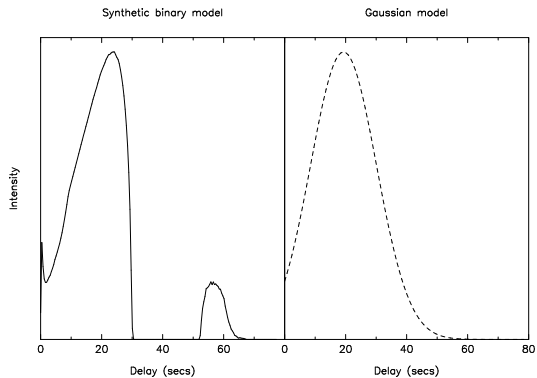}
\caption{A comparison of the best fit transfer functions for GRO~J1655-40
from our two modeling methods. On the left is the synthetic X-ray binary
model transfer function and on the right is the acausal Gaussian transfer
function.}
\label{j1655modelvsgauss}
\end{center}
\end{figure}

The transfer functions created by our X-ray binary model, using the binary
parameters determined by \scite{orosz97} and summarized in
Table~\ref{parstable} were used to create synthetic \HST\ lightcurves by
convolving them with the \RXTE\ lightcurves. The badness of fit between
this synthetic reprocessed lightcurve and the real reprocessed lightcurve,
from the \HST\ observations was calculated. The fitting again contained 3
parameters, which were optimized as before. The best-fit parameters,
together with their 1-parameter 1-sigma confidence regions are given in
Table~\ref{j1655modeltable}. We find that the disc extends to 67 percent
of the distance to the inner Lagrangian point, with an opening angle of 14
degrees and an exponent of 3.8. 

\subsubsection{comparison of the two models}

This best-fit solution from our modeling is shown along with the best-fit
solution from the acausal Gaussian fitting in
Figure~\ref{j1655modelvsgauss}. The best-fit solution for this modeling
shows a similar $\chi^2$ to the Gaussian fitting, which shows that our
modeling of the binary system and simple Gaussian fits are both good fits
to the data. The first and second moments from the model transfer function
are, as expected, similar to the best fit parameters from the Gaussian
fitting. The opening angle for the disc from our synthetic transfer functions, whilst being surprisingly large is comparable with that derived by \scite{orosz97} from lightcurve fitting shortly after the previous outburst in March 1995. They found that the opening angle of the disc was 11\degree (see Table~6 from \scite{orosz97}) compared to our value of 14\degree. 

%
%
\section{discussion}
\label{discuss}

We have used the correlated X-ray and optical variability seen in Low Mass
X-ray Binaries to determine geometric parameters for the binary system \novasco. These parameters are principally the size and shape of
the accretion disk. This is inferred from the relative contributions to
the time delay transfer function of the different regions of the binary. 
We have used time delay transfer functions, along with the known binary
parameters to find best fit solutions to the data along with their
corresponding confidence regions. 

There is evidence for a larger fraction of disk
reprocessing. The geometric parameters determined from our fit give an
opening angle of 19 degrees and a very flared geometric shape, implying
that most of the reprocessing is taking place in the outer regions of the
disc. This means that the companion star is almost entirely shielded from
X-rays, which in turn reduces the mass accretion rate, which is driven by
irradiation. The observations of \novasco\ took place during outburst,
which could explain the flared shape of the outer disk. Another possible
interpretation is that there is a localized region of enhanced
reprocessing in the outer disk that is non-axisymmetric that is adding a
large component from the outer disc region to the transfer function but is
difficult to distinguish from reprocessing from a thick disk. This would
explain the high value of $\beta$ observed, as the model attempts to move
all the reprocessing to the outer disk. The most likely site for this
reprocessing would be at the disk-stream interaction point, where the rim
of the disk swells greatly. This inhomogeneity is observed in \novasco\ as
X-ray dips around binary phase 0.8. This is the site proposed for the
reprocessing of the X-ray bursts seen in 4U/MXB~1636-53 \cite{matsouka84}. 

The effect of reprocessing timescales in different regions of the binary
also needs to be studied in more detail, with detailed radiative transfer
models. The incidence angle of the X-rays to the atmosphere may cause
large variations in the timescale. Normally incident X-ray photons may be
reprocessed deeper within the companion star and hence take longer to
diffuse to the surface, than those with a grazing incidence angle. This
would also affect the ratio of disk to companion star reprocessing, as the
incidence angles for the disk will be predominantly grazing.

In order to distinguish between these scenarios and determine the
importance of reprocessing timescales it is necessary to have data with
better phase coverage to observe a change in the time delay of this
enhanced region as a function of binary phase and any reprocessing
timescale effects. It is clear from our analysis that the companion star
is responsible for a small fraction of the instantaneously reprocessed
flux, see figure~\ref{j1655modelvsgauss}. 
 
This is the first data showing correlated X-ray and optical variability
with sufficient time resolution and a long enough base-line to do this
form of echo-tomography. We have used the time delayed optical variability
observed from \novasco\ to constrain the binary parameters of
these systems and find that both the accretion disk and companion star can
be important regions for reprocessing in XRBs. 

%
%
\section*{Acknowledgments}
KOB was supported by a PPARC Research Studentship during much of this work. Support for this work was also provided by NASA through grant numbers GO-6017-01-94A from the Space Telescope Science Institute, which is operated by the Association of Universities for Research in Astronomy, Incorporated, under NASA contract NAS5-26555 and NAS5-32490 for the \RXTE\ project.  This work made use of the \RXTE\ Science Center at the NASA Goddard Space Flight Center and has benefitted from the NASA Astrophysics Data System Abstract Service. We would also like to thank the anonymous referee for many useful comments and suggestions.
%
%
\bibliographystyle{mn}
\bibliography{/home/kso/Papers/ksobib}
%
%
\label{lastpage}
%
\end{document}